\documentclass[aps,prl,twocolumn,,superscriptaddress,floatfix]{revtex4-1}
\usepackage[colorlinks=true,allcolors=blue, bookmarks=false]{hyperref}
\usepackage{amsfonts,amsmath,amssymb}
\usepackage{graphicx}

\DeclareMathOperator{\Tr}{Tr}

\begin{document}

\title{Non-ergodicity in the Anisotropic Dicke model}

\author{Wouter Buijsman} 

\email{w.buijsman@uva.nl}

\author{Vladimir Gritsev}

\affiliation{Institute for Theoretical Physics Amsterdam and Delta Institute for Theoretical Physics, University of Amsterdam, Science Park 904, 1098 XH Amsterdam, The Netherlands}

\author{Rudolf Sprik}

\affiliation{Van der Waals-Zeeman Institute, University of Amsterdam, Science Park 904, 1098 XH Amsterdam, The Netherlands}

\date{\today}

\begin{abstract}
We study the ergodic -- non-ergodic transition in a generalized Dicke model with independent co- and counter rotating light-matter coupling terms. By studying level statistics, the average ratio of consecutive level spacings, and the quantum butterfly effect (out-of-time correlation) as a dynamical probe, we show that the ergodic -- non-ergodic transition in the Dicke model is a consequence of the proximity to the integrable limit of the model when one of the couplings is set to zero. This can be interpreted as a hint for the existence of a quantum analogue of the classical Kolmogorov-Arnold-Moser theorem. Besides, we show that there is no intrinsic relation between the ergodic  -- non-ergodic transition and the precursors of the normal -- superradiant quantum phase transition.
\end{abstract}

\maketitle

Non-ergodic quantum dynamics and, more generally, non-ergodic phases of quantum many-body systems have recently attracted a great interest in the condensed matter community \cite{Rigol09, Pal10, Bertini15, Giraud16}. One of the paradigmatic subclasses of these systems is provided by integrable models of quantum many-body systems, such as the anisotropic Heisenberg (XXZ) spin-$1/2$ chain \cite{Orbach58}, the one-dimentional Hubbard model \cite{Essler05}, central spin models \cite{Khaetskii02}, as well as various other interacting one-dimensional bosonic \cite{Cazalilla11} or fermionic \cite{Giamarchi03} models. While integrable systems are characterized by an infinite number of integrals of motion leading to non-ergodic dynamics, they do not exhaust all possible non-ergodic phases. In fact, many-body localized systems represent a new class of non-ergodic phases with the ergodic -- non-ergodic transition (ENET) driven by disorder strength \cite{Pal10, Imbrie16}. For these systems, dynamically emergent conserved quantities are responsible for a variety of distinctive properties of many-body localized phases. Non-ergodic phases could also exist in driven and dissipative quantum systems \cite{Bagnoli03, Drotos16}. The existence of non-ergodic phases breaking traditional statistical physics by not satisfying the eigenstate thermalization hypothesis \cite{Srednicki94} can possibly be linked with the existence of  a yet unknown quantum version of the classical Kolmogorov-Arnold-Moser theorem \cite{Kolmogorov54, Arnold89} (qualitatively stating that classical integrable systems remain quasi-integrable under weak perturbations). Despite several attempts to identify such a quantum theorem \cite{Hose83, Barmettler13, Bertini15, Brandino15}, fundamental questions are still open, and progress mostly lies in the observation of indirect signatures in specific models.

Here, we discuss the emergence of extended non-ergodic phases in a generalized version of the Dicke model \cite{Dicke54}. This model has two independent light-matter coupling constants, corresponding to the co- and counter-rotating terms in the Hamiltonian. The model can be derived as an effective model starting from three- or four-level emitter schemes \cite{Dimer07, Tomka15}. While it is Bethe ansatz integrable when one of the couplings is zero (then representing a variant of the Gaudin model \cite{Gaudin76}), tuning the coupling parameter from a nonzero value and considering it as a perturbation allows to study the transition from the non-ergodic phase (corresponding to quasi-integrability) to the ergodic phase (associated with quantum chaotic behavior). We show that this transition occurs at finite values of the integrability-breaking parameter, with the non-ergodic phase occupying an extended region of the phase diagram. Interestingly, this effect can be observed experimentally \cite{Baumann10, Baden14}. Second, we show that there is a clear difference between the ENET and the precursors of the normal -- superradiant quantum phase transition \cite{Wang73}, thereby shining a new light on the question whether ENETs and normal -- superradiant quantum phase transitions can be intrinsically related \cite{Emary03, *Emary03-2, Perez11, Lobez16}. 

The Dicke model is a paradigmatic model to benchmark tools detecting quantum chaos \cite{Song09, Giorda10, Wisniacki10, Lobez16}. Here, we use several complementary methods to detect the ENET: we study the level statistics, the average ratio of consecutive level spacings, and the quantum butterfly effect. All these methods are complimentary, while indicating the same shape of the ENET. We note that in these and many other studies in the past, signatures of non-ergodic behaviour were quantified by quantities related to eigenvalues and eigenfunctions. However, in the present many-body context we emphasize the need of \emph{dynamical} probes of non-ergodicity. The quantum butterfly effect (also known as `scrambling' or `out-of-time correlation') \cite{Shenker14, Shenker15, Hosur16, Fu16, Roberts16} serves this aim for us. This recently developed tool has been used for example in quantum gravity \cite{Roberts16}, black hole physics \cite{Shenker14} and many-body localization \cite{Chen16}. To our knowledge, its use in quantifying the phase diagram of a quantum optical system is new.

\textit{Anisotropic Dicke model}.--- 
We consider the Anisotropic Dicke model (ADM)
\begin{align} 
\begin{split}
H & = \omega a^\dagger a + \omega_0 J_z  + \frac{g_1}{\sqrt{2j}} \left( a^\dagger J_- + a J_+ \right) \\
& + \frac{g_2}{\sqrt{2j}} \left( a^\dagger J_+ + a J_- \right),
\end{split} 
\label{eq: H}
\end{align}
where $a$, $a^\dagger$  are bosonic (cavity mode) operators satisfying $[a,a^\dagger] = 1$ in units $\hbar = 1$ and  $J_{\pm,z}=\sum_{i=1}^{2j} \frac{1}{2} \sigma^{(i)}_{\pm,z}$ are angular momentum operators of a pseudospin with length $j$ composed of $N=2j$ non-interacting spin-$1/2$ atoms described by the Pauli matrices $\sigma_{\pm, z}^{(i)}$ acting on site $i$. In the following, we work in the basis $\{ |n \rangle \otimes |j, m \rangle \}$ with $a^\dagger a |n \rangle = n |n \rangle$ and $J_z |j, m \rangle = m |j, m \rangle$. The ADM describes the interaction between a single-mode bosonic field with frequency $\omega$ and the atoms with level splitting $\omega_0$, within the dipole approximation coupled to the field with coupling parameters $g_1$ and $g_2$ for the co- and counter rotating terms, respectively. Several experimental realizations of the ADM have been proposed \cite{Dimer07, Baden14, Zou14, Song16}. For $g_1 = g_2 = g$, the ADM reduces to the Dicke model with coupling parameter $g$. The ADM possess a parity symmetry $[H, \Pi] = 0$ with $\Pi = \exp(i \pi [a^\dagger a + J_z + j])$ having eigenvalues $\pm 1$. Here, the focus is restricted to the positive parity subspace, which includes the ground state for the parameter ranges considered in this Letter (verified numerically). When applying the rotating-wave approximation, i.e. setting $g_2 = 0$, the total number of excitations $n + m + j$ is conserved, making the ADM Bethe ansatz integrable \cite{Gaudin76}. By rotating $J_y \to - J_y$, $J_z \to - J_z$ and setting $\omega_0 \to - \omega_0$, the ADM with $g_1 = 0$ maps onto the ADM with $g_2 = 0$, showing that the ADM is integrable for $g_1 = 0$ or $g_2 = 0$. In the thermodynamic limit $j \to \infty$, the ADM exhibits a second-order quantum phase transition \cite{Hioe73} at $g_1 + g_2 = \sqrt{\omega \omega_0}$ with order parameter $a^\dagger a / j$, separating the normal phase at $g_1 + g_2 < \sqrt{\omega \omega_0}$ with $\langle a^\dagger a\rangle / j = 0$ from the superradiant phase with $\langle a^\dagger a \rangle / j = \mathcal{O}(1)$. For finite $j$, it has been shown numerically that the Dicke model displays a transition from non-ergodic to ergodic behaviour with an increasing value of $g$ at $g \approx \sqrt{\omega \omega_0} / 2$, which is believed to be caused by the precursors of the quantum phase transition \cite{Emary03, *Emary03-2}. Both in the quantum and semiclassical regime, this transition has been investigated extensively \cite{Perez11, Altand12, Altand12-2, Brandes13, Chavez16}.

\begin{figure}
\includegraphics[width=\columnwidth]{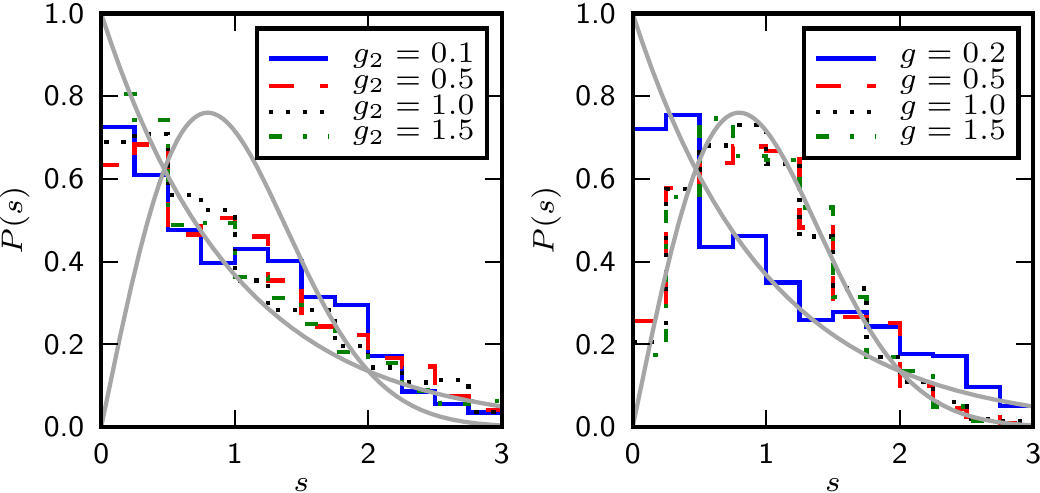} 
\caption{The normalized distribution of level spacings $s_n = E_{n+1} - E_n$ for the ADM with $\omega = \omega_0 = 1$ and $j=10$ at $g_1 = 0.1$ (left) and $g = g_1 = g_2$ (right). The histograms are drawn from the sorted energy levels $E_n$ with $n$ ranging from $200$ to $1000$, where the lowest levels are left out to account for the non-uniform density of states. As a reference, the Poissonian and Wigner-Dyson distributions are shown in gray.}
\label{fig: spacing}
\end{figure}

\textit{Level statistics}.---
The onset of ergodic behaviour is typically diagnosed by inspection of the level spacing distribution \cite{Stockmann99}. Let $\{ E_n \}$ denote the energy levels of the ADM in ascending order. Under the assumption that the density of states equals unity, the distribution $P(s)$ of the level spacings $s_n = E_{n+1} - E_n$ is given by the Poissonian distribution $P(s) = \exp(-s)$ for non-ergodic systems following the Berry-Tabor conjecture \cite{Berry77} and the Wigner-Dyson distribution $P(s) = \frac{\pi}{2} s \exp(- \frac{\pi}{4} s^2)$ for ergodic systems invariant under orthogonal transformations satisfying the Bohigas-Giannoni-Schmit conjecture \cite{Bohigas84}. Fig. \ref{fig: spacing} shows the level spacing distribution for the ADM with $\omega = \omega_0$ and $j=10$ for several values $g_{1,2}$ obtained by exact diagonalization. The ADM is integrable and hence non-ergodic \cite{Srednicki94} at $g_1 = 0$ or $g_2 = 0$. One observes that there is an ENET when following the line $g_1 = g_2$, whereas the system remains non-ergodic along a line close to the integrable limit $g_1 = 0$, strongly suggesting that the ENET is a consequence of the integrability at $g_1 = 0$ or $g_2 = 0$.

\begin{figure}
\includegraphics[width=\columnwidth]{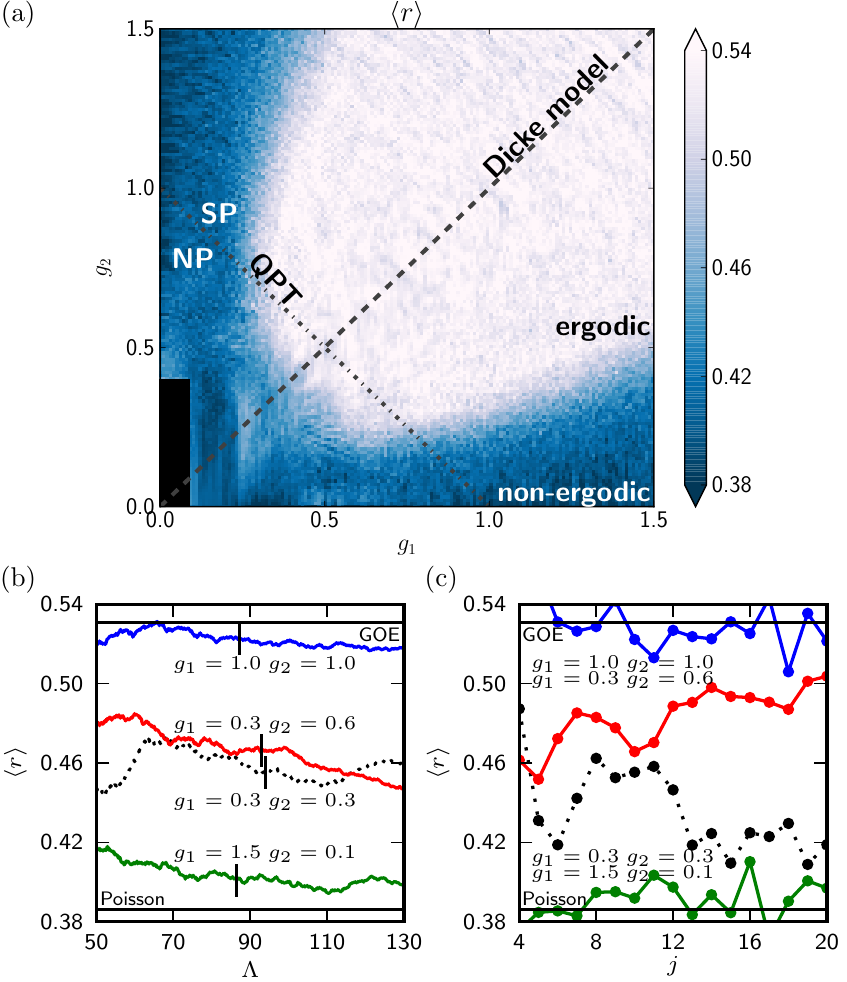} 
\caption{(a) The average $\langle r \rangle$ of $r_n$ taken over the lowest $1000$ energy levels of the ADM with $\omega = \omega_0 = 1$ and $j=10$ as a function of $g_{1,2}$. The Dicke model and the quantum phase transition (QPT) between the normal (NP) and superradiant (SP) phases are indicated by a dashed and a dash-dotted line, respectively. The lower left corner, where the data artificially suggests non-ergodic behaviour (see main text), has been masked. (b) The dependence of $\langle r \rangle$ on the upper energy window cutoff $\Lambda$ for various values $g_{1,2}$. The cutoffs for which the energy windows contain the lowest 1000 energy levels are indicated by black lines. (c) The dependence of $\langle r \rangle$ on $j$ for various values $g_{1,2}$.}
\label{fig: ratio}
\end{figure}

\textit{Average ratio of consecutive level spacings}.---
Aiming to provide a more complete view on where the ENET occurs, we study the average $\langle r \rangle$ over $n$ of the ratio of consecutive level spacings 
 \begin{equation}
 r_n = \min \left( \frac{s_n}{s_{n-1}}, \frac{s_{n-1}}{s_n} \right),
 \label{eq: ratiodef}
 \end{equation}
which is independent of the local density of states, and can be used to localize the transition from ergodicity to non-ergodicity \cite{Atas13}. The average $\langle r \rangle$ takes a value $\langle r \rangle = 2 \ln 2 - 1 \approx 0.386$ for Hamiltonians from the Poissonian ensemble corresponding to the class of non-ergodic systems as described above, or a value $\langle r \rangle = 0.5307(1)$ for Hamiltonians from the Gaussian orthogonal ensemble (GOE), corresponding to the above discussed class of ergodic systems. Fig. \ref{fig: ratio} shows $\langle r \rangle$ for the ADM with $\omega = \omega_0 = 1$ and $j=10$ as a function of $g_{1,2}$. Clearly, the ENET along the line $g_1 = g_2 = g$ is caused by the integrability of the ADM for $g_1 = 0$ or $g_2 = 0$, and is not related to the precursors of the quantum phase transition at $g_1 + g_2 = 1$ (verified numerically to be close to this line) as it extends over the full ranges $g_1=0$ and $g_2=0$. One observes that the width of the non-ergodic regions on the lower and left sides of the plot increases with increasing values of the coupling parameters, which we expect to be a result of the integrability of the Dicke model in the limit $g / \omega_0 \to \infty$ obtained by rotating $J_x \to J_z$, $J_z \to J_x$ and translating $a^\dagger \to a^\dagger - 2 m g / \sqrt{2j}$, where $\omega_0 J_x$ is treated perturbatively. Except for the region around $g_1 = g_2 \approx 0.3$, there are no qualitative differences when varying the system size or number of energy levels taken into account. For large values of $j$, the value of $\langle r \rangle$ converges to either the limiting value for ergodic or non-ergodic systems, depending on the values of $g_{1,2}$. Expanded up to first order in $g_{1,2}$ around $g_1 = g_2 = 0$, the states $|n, j, m \rangle$ and $|n \pm 1, j, m \mp 1  \rangle$ are near-degenerate with energies for $\omega = \omega_0$ given by $\omega (n+m)$ and $\omega (n+m) \pm g_1 \sqrt{j + j^2 + m - m^2 + 2(j+j^2 - m^2)n} / \sqrt{2j}$, respectively. This clustering phenomenon with equally separated energy levels leads to a large value of $\langle r \rangle$, artificially suggesting strong non-ergodic behavior near $g_1 = g_2 = 0$.

\textit{Quantum butterfly effect}.---
The connection between level statistics and ergodicity is not absolute, and counterexamples do exist \cite{Berry77, Finkel05}. Here, we utilize the quantum butterfly effect \cite{Roberts15} as an independent dynamical tool to validate the above result. Let $V(t)$ and $W(t)$ denote time-evolving Hermitian operators for which $[V(0), W(0)] = 0$ and $[W(0), H] \neq 0$. In an ergodic phase, one expects a small perturbation by applying $V$ at time $t=0$ to strongly affect the outcome of a later measurement of $W$, thereby contrasting with a non-ergodic phase. This effect can be measured by the degree of non-commutativity
\begin{equation}
F(t) = \frac{1}{2} \left( \langle V^\dagger(0) W^\dagger(t) V(0) W(t) \rangle_\beta + \text{h.c.} \right)
\label{eq: F}
\end{equation}
written in the Heisenberg picture, with $\langle \mathcal{O} \rangle_\beta$ denoting a thermal average at inverse temperature $\beta = 1/T$ in units $k_b = 1$ for an operator $\mathcal{O}$ given by $\langle \mathcal{O} \rangle_\beta = \Tr( \mathcal{O} e^{- \beta H}) / \Tr(e^{- \beta H})$. Considering, for the moment, the pure state $| \Psi \rangle$ at $t=0$, this effect can be understood by viewing the first term in $F$ (a similar argument holds for the second term) as the overlap between the states $| \Psi_1 \rangle = W(t) V(0)|\Psi \rangle$ and $| \Psi_2 \rangle = V(0) W(t)| \Psi \rangle$. Since $[V(0),W(0)] = 0$, $|\Psi_{1,2} \rangle$ initially fully overlap. With evolving time, the overlap will decrease to a value depending on the size of the accessible Hilbert space in an ergodic phase, thereby contrasting with a non-ergodic phase, provided that the perturbation is small. Considering a thermal average, it is expected that $F(t)$ eventualy approaches a constant value. Hence, $F(t)$ takes a relatively small or large value in ergodic and non-ergodic phases, respectively. 
\begin{figure}
\includegraphics[width=\columnwidth]{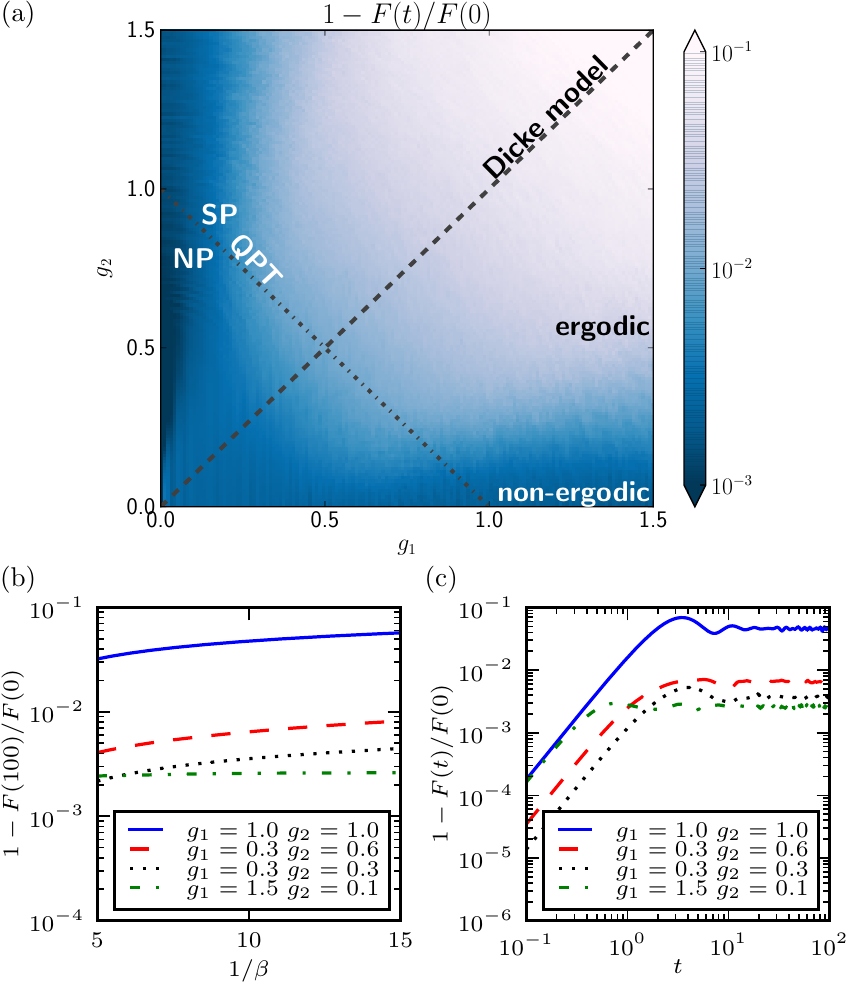} 
\caption{(a) Density plot of $1-F(t)/F(0)$ at $t=100$ and $\beta = 1/10$ for the ADM with $\omega = \omega_0 = 1$ and $j=5$. Annotations are similar to Fig. \ref{fig: ratio}. Compared to Fig. \ref{fig: ratio}, $j$ is chosen smaller to account for the computational cost of evolving $W(t)$ over time. (b) The temperature dependence of $1-F(100)/F(0)$ for various values $g_{1,2}$. (c) The dependence of $1-F(t)/F(0)$ on time for various values $g_{1,2}$.}\label{fig: butterfly}
\end{figure}
The quantum butterfly effect can -- in principle -- be measured experimentally \cite{Swingle16, Zhu16}. By adding a probe and control qubit to the system, $| \Psi \rangle$ can be duplicated, after which separate states $W(t) V(0) | \Psi \rangle$ and $V(0) W(t) | \Psi \rangle$ can be obtained by the proper use of entangling gates \cite{Pedernales14}. Subsequently, the value of $F(t)$ can be obtained by measuring the overlap of these states.

In measuring the quantum butterfly effect in the ADM, we take $W=V$ with $V = a^\dagger a + 100$ in the thermal ensemble at inverse temperature $\beta$. For the parameters under consideration, the number of bosonic excitations is small compared to $100$ (verified numerically), such that $V$ is close to the scaled unit operator, keeping the perturbation small. Fig. \ref{fig: butterfly} shows $1-F(t)/F(0)$ for the ADM after equilibration as a function of $g_{1,2}$. This quantity is relatively large (small) in an ergodic (non-ergodic) phase. The Dicke model displays a quasi-integrable structure at low energies \cite{Perez11}, which is characterized by e.g. a Poissonian level spacing distribution. Even though part of active research \cite{Brandes13,Chavez16} the origin of this phenomenon is still unclear. Here, we choose $\beta = 1/10$ for which the this low-energy regime does not qualitatively influence the results (verified numerically). One observes qualitative agreement with Fig. \ref{fig: ratio}, showing that the ENET in the ADM is a consequence of the integrability at $g_1 = 0$ or $g_2 = 0$. The result is qualitative independent of variations in the temperature or equilibration time for the wave function to spread out over the full accessible Hilbert space. We also confirm that the ENET is unrelated to the precursors of the normal -- superradiant quantum phase transition.

\begin{figure}
\includegraphics[width=\columnwidth]{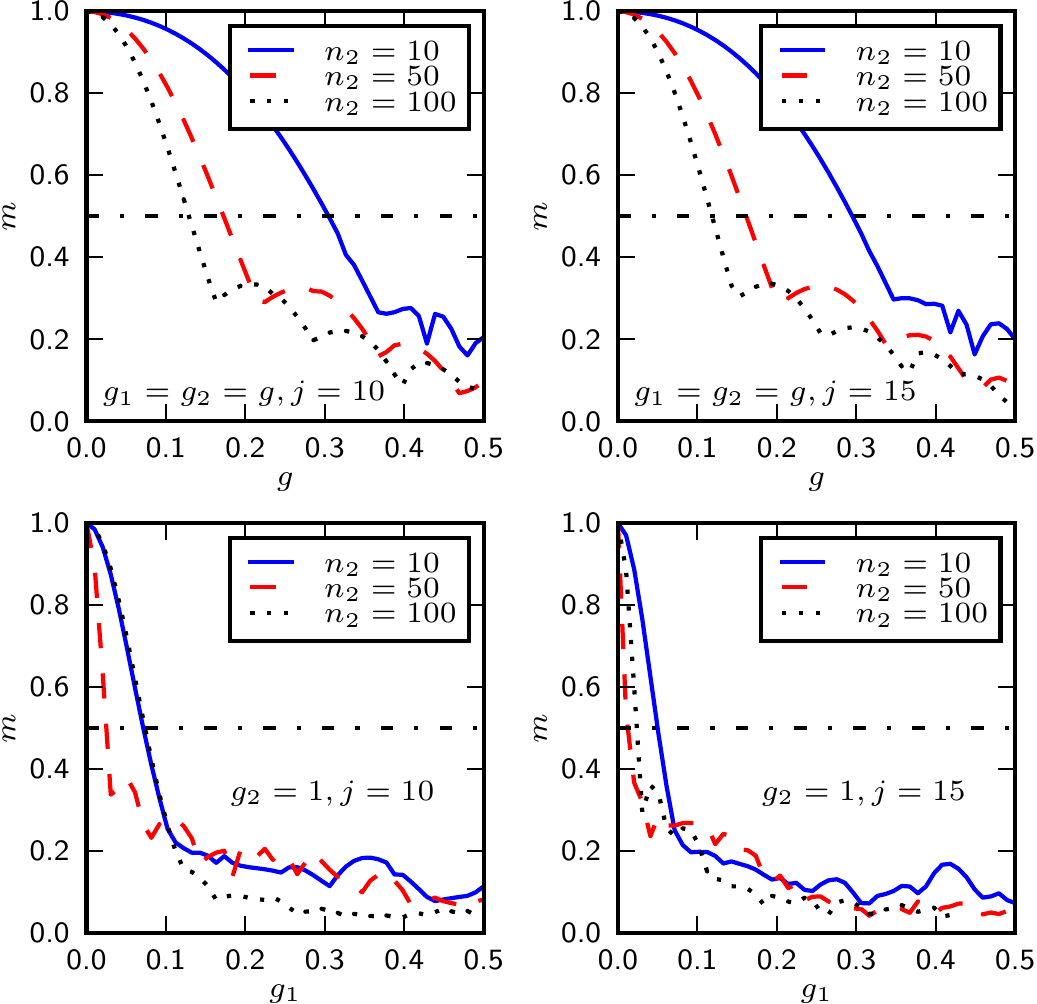} 
\caption{The quantity $m$ as defined in the main text for the ADM along the line $g_1 = g_2 = g$ ($g_1 = 1$) with $g=0$ ($g_2 = 0$) denoting the integrable Hamiltonian in the upper (lower) part. The results are shown for system size $j=10$ (left) and $j=15$ (right).} \label{fig: HoseTaylor}
\end{figure}

\textit{Discussion}.--- 
Besides the diagnostics for ergodicity utilized in this Letter, multiple alternative measures exist. Let $H(\lambda)$ denote a parameter-dependent Hamiltonian, and suppose $H(\lambda_1)$ is integrable, opposed to $H(\lambda_2)$. Ref. \cite{Hose83} argues that $H(\lambda_2)$ is non-ergodic if for any eigenstate $|n_2 \rangle$ there is an eigenstate $| n_1 \rangle$ of $H(\lambda_1)$ such that $| \langle n_1 | n_2 \rangle|^2 > 1/2$. Here, we investigate this proposal by determining the quantity $m = \max_i | \langle i | n_2 \rangle|^2$ with $| i \rangle$ running over all eigenstates of the integrable Hamiltonian, and $|n_2 \rangle$ denoting the $n_2$-th eigenstate of the non-integrable Hamiltonian labeled according to the corresponding eigenvalues in ascending order. Fig. \ref{fig: HoseTaylor} shows $m$ for the ADM with $g_1 = g_2 = 0$ (upper part) and $g_1 =  0$, $g_2 =1$ (lower part) as the integrable Hamiltonian for several eigenvectors and system sizes as a function of $g = g_1 = g_2$ and $g_1$, respectively. Opposed to the results above, this measure suggests that the ENET moves towards the integrable regime $g_1 = 0$ or $g_2 = 0$ with an increasing energy scale or system size $j$. We believe that future studies focusing on this decrepancy would be helpful.

\textit{Conclusions}.--- 
We have shown that the ergodic -- non-ergodic transition in the Dicke model is a result of integrability of the Anisotropic Dicke model when setting one of the coupling constants to zero. We have shown that there is an extended non-ergodic region, which can be considered as a hint for the existence of a still elusive quantum version of the Kolmogorov-Arnold-Moser theorem. Similar observations have been made for e.g. Gaudin models \cite{Barmettler13}, spinless fermion models \cite{Bertini15}, or one-dimensional Bose gases \cite{Brandino15}. Experimental setups feasible to verify the results experimentally have been proposed \cite{Dimer07}. We have used both the level spacing distribution and the average ratio of consecutive level spacings as static and the quantum butterfly effect as dynamical probes for ergodicity. Besides, we have shown that there is no intrinsic relation between the ergodic -- non-ergodic transition and the precursors of the normal -- superradiant quantum phase transition. We expect that a similar approach as used in this Letter can be used to find extended non-ergodic phases in other quantum many-body systems, such as disordered spin chains.

\textit{Acknowledgements}.--- 
The work of W. B. and V. G. is part of the Delta-ITP consortium, a program of the Netherlands Organization for Scientific Research (NWO) that is funded by the Dutch Ministry of Education, Culture and Science (OCW).

\bibliography{references}

\end{document}